\documentclass[useAMS,usenatbib,onecolumn]{mn2e}

\newcommand{\kbf}       {\mbox{\boldmath$k$}}

\newcommand{\fbf}       {\mbox{\boldmath$f$}}
\newcommand{\rbf}       {\mbox{\boldmath$r$}}

\newcommand{\vbf}       {\mbox{\boldmath$v$}}

\newcommand{\omegabf}       {\mbox{\boldmath$\omega$}}

\newcommand{\sigmabf}       {\mbox{\boldmath$\sigma$}}
\newcommand{\Omegabf}       {\mbox{\boldmath$\Omega$}}

\def\rads{\hbox{\rm\hskip.35em  rad s}$^{-1}$}
\def\gcc{\hbox{\rm\hskip.35em  g cm}$^{-3}$}
\def\cms{\hbox{\rm\hskip.35em  cm s}$^{-1}$}

\def\lap{\hbox{${_{\displaystyle<}\atop^{\displaystyle\sim}}$}}
\def\gap{\hbox{${_{\displaystyle>}\atop^{\displaystyle\sim}}$}}

\usepackage{graphicx}
\usepackage{subfigure}

\title[Instability of Superfluid Flow]
{Instability of Superfluid Flow in the Neutron Star Inner Crust}
\author[B. Link]{B. Link\thanks{E-mail:
link@physics.montana.edu;}\\
Department of Physics, Montana State University, Bozeman, MT
59717, USA}
\begin{document}


\pagerange{\pageref{firstpage}--\pageref{lastpage}} \pubyear{2011}

\maketitle

\label{firstpage}

\begin{abstract}
Pinning of superfluid vortices to the nuclear lattice of the inner
crust of a neutron star supports a velocity difference between the
superfluid and the solid as the star spins down. Under the Magnus
force that arises on the vortex lattice, vortices undergo {\em vortex
creep} through thermal activation or quantum tunneling.  We examine
the hydrodynamic stability of this situation. Vortex creep introduces
two low-frequency modes, one of which is unstable above a critical
wavenumber for any non-zero flow velocity of the superfluid with
respect to the solid. For typical pinning parameters of the inner
crust, the superfluid flow is unstable over length scales $\lap 10$ m 
and over timescales as fast as months. The vortex lattice could degenerate into
a tangle, and the superfluid flow could become turbulent. Unexpectedly
large dissipation would suppress this instability.
\end{abstract}


\section{Introduction}
\maketitle


The dynamics of the superfluid interior of a neutron star is central
to understanding a variety of phenomena that includes observed spin
glitches, stochastic spin variations and thermal evolution, as well as
possible precession and r-modes.  In this connection, the possible
importance of hydrodynamic instabilities in neutron stars has 
become a question of considerable
interest. \citet{peralta_etal05,peralta_etal06} have shown that
differential rotation in the core, resulting from a spin glitch or
possibly causing it, drives an Ekman flow along the rotation axis than
can excite a variant of the ``Glaberson-Donnelly'' counterflow
instability in liquid helium \citep{Glaberson_etal74}; transitions
between laminar flow and fully-developed turbulence could drive spin
glitches. This instability could also be excited in precessing neutron
stars \citep{gaj08a,vl08}. Unstable shear layers \citep{pm09} and
r-mode instabilities \citep{ga09} in the outer core may also play a
role in glitches.

From the standpoint of building a realistic theory of neutron star
seismology with which to interpret observations, it is important to
identify hydrodynamic instabilities of possible relevance.  The
possibility of turbulent instabilities in the neutron star inner
crust, the region from the neutron drip density to about half nuclear
saturation density, has received little attention in this regard. Here
the vortices that thread the rotating superfluid are predicted to
interact with nuclei with energies of $\sim 1-5$ MeV per nucleus in
the denser regions
\citep{alpar77,eb88,dp06,abbv07}. Recent work has shown 
that this interaction will pin vortices to nuclei, regardless of the
details of the pinning potential \citep{link09}.  As the star spins
down, the differential velocity between the superfluid and the pinned
vortices approaches the critical value at which the hydrodynamic lift
force on vortices, the {\em Magnus force}, would unpin them. As
suggested long ago by \citet{ai75}, the spin glitches seen in neutron
stars could arise from large-scale vortex unpinning from nuclei,
wherein the threshold for pinning is exceeded.  Below the critical
velocity, pinned vortices slowly creep through thermal activation
\citep{alpar_etal84,leb93} or quantum tunneling \citep{leb93}, driven
by the Magnus force. Here we demonstrate the existence of a
hydrodynamic instability related to the vortex creep process that
could grow over timescales as short as months. 

In the next section, we describe vortex pinning in the inner crust. In
\S 3 we give the stability analysis. In \S 4, we discuss
hydrodynamic wave solutions in the case of no background flow. In \S
5, we describe the  hydrodynamic instability that arises when
vortices move slowly through the nuclear lattice. In \S 6, we
calculate the vortex mobility, which we apply to obtain the growth
rate of the instability. We conclude with a discussion of the 
possibility that the inner-crust superfluid becomes turbulent. 

\section{Vortex Pinning}

First we calculate the critical velocity between the superfluid and the
crust that can be sustained by vortex pinning. 
We will use these results in \S 6 to calculate the vortex
mobility and the growth rate of the instability. 

Vortex pinning fixes the local superfluid velocity in the laboratory
frame. As the crust spins down, a velocity difference $v$ between the
pinned vortices and the superfluid develops. The Magnus force per unit
length of vortex is
\begin{equation}
f_{\rm mag}=\rho\kappa v,
\end{equation}
where $\rho$ is the superfluid mass density, $\kappa\equiv h/m$ is the
quantum of vorticity, and $m$ is twice the neutron mass.  Let $F_p$ be
the characteristic force of the vortex-nucleus potential. Above a
critical velocity difference $v_c$, the Magnus force will exceed the
pinning force, and vortex pinning is not possible.  If a vortex could
bend to intersect nuclei of average spacing $a$, the critical velocity
difference $v_c$ would be given by
\begin{equation}
\rho\kappa v_ca=F_p.
\end{equation}

A vortex has a large self energy (tension) that typically prevents it
from bending over a length scale $a$.  If the tension were infinite, a
vortex could not pin at all, since the vortex would remain straight
and the forces from nuclei that surround the vortex would cancel on
average. For finite tension, the vortex can bend over a length
$l_p>a$, and the critical velocity is given instead by
\begin{equation}
\rho\kappa v_c l_p=F_p, 
\end{equation}
giving a critical velocity
\begin{equation}
v_c =\frac{F_p}{\rho\kappa a}\left(\frac{a}{l_p}\right). 
\label{vc0}
\end{equation}
Tension lowers the critical velocity by a factor $a/l_p$. 
To calculate $l_p$, let $\rbf_v(z)$ be a vector in the $x-y$ plane
that gives the shape of the pinned vortex.  The energy of a static vortex in a
pinning field $V(\rbf_v)$, in the absence of an ambient superfluid
flow, is
\begin{equation}
E_v=\int dz\, \left(\frac{1}{2}T_v\left|\frac{d\rbf_v(z)}{dz}\right|^2 +
V(\rbf_v)\right), 
\label{e}
\end{equation}
where $T_v$ is vortex tension,
typically 1 MeV fm$^{-1}$. On average, over a length $l_p$ 
the vortex bends by an amount $\delta r_v$ to intersect one nucleus in
a volume $l_p\pi (\delta r_v)^2$.  The quantities $l_p$ and $\delta r_v$
are therefore related by
\begin{equation}
a^{-3} l_p \pi (\delta r_v)^2=1.
\label{mfp}
\end{equation}
The energy of the vortex per unit length, from eq. (\ref{e}), 
is approximately 
\begin{equation}
\frac{E_v}{l_p}\simeq \frac{1}{2}T_v\frac{(\delta r_v)^2}{l_p^2}
-\frac{E_p}{l_p}, 
\end{equation}
where $E_p$ is the interaction energy between a vortex and a single
nucleus, typically $\sim 1$ MeV. 
Contributions to the potential by nuclei that the vortex does
not intersect have been ignored; these contributions will largely
cancel.  Minimization of $E_v/l_p$ with respect to $l_p$, using
eq. (\ref{mfp}), gives
\begin{equation}
\frac{l_p}{a}=\left(\frac{3aT_v}{2\pi E_p}\right)^{1/2}.
\label{lp}
\end{equation}
The vortex tension $T_v$ is due mainly to the kinetic energy per unit length
of vortex due to circulation abut the vortex, and 
takes the form \citep{thomson1880,fetter67},
\begin{equation}
T_v=\frac{\rho\kappa^2}{4\pi}(0.116-\ln k_v\xi), 
\label{tension}
\end{equation}
where $\xi$ is the
radius of the vortex core and $k_v$ is the characteristic bending
wavenumber, $k_v=\pi/2l_p$.  
 
For typical conditions of the inner crust, the ratio $l_p/a$ is much
larger than unity.  At a density $\rho=5\times 10^{13}$ \gcc\, the
lattice spacing is $a\simeq 50$ fm and the radius of the vortex core
is $\xi\simeq 10$ fm. For $E_p=1$ MeV, simultaneous solution of
eqs. (\ref{lp}) and (\ref{tension}) gives $l_p\simeq 9a$. The ratio
$l_p/a$ increases for weaker pinning. For example, for $E_p=0.1$ MeV,
the pinning length becomes $l_p\simeq 32a$. For $E_p=10$ MeV,
unrealistically large according to recent calculations, $l_p=2a$.

Combining eq. (\ref{lp}) with eq. (\ref{vc0}) gives the critical velocity,
modified by vortex tension, 
\begin{equation}
v_c= \frac{F_p}{\rho\kappa
a}\left(\frac{a}{l_p}\right)=
\frac{E_p}{\rho\kappa a\xi}
\left(\frac{2E_p}{3aT_v}\right)^{1/2},
\label{vcrit}
\end{equation}
where we have taken $F_p=E_p/\xi$. Eq. (\ref{vcrit}) was 
found in the numerical simulations of the
dynamics of an isolated vortex in a random potential \citep{link09}.
This equation shows that pinning is weakened by vortex tension. 
For $E_p=$ 1 MeV and $\xi=10$ fm, the critical velocity is $v_c\simeq
4\times 10^5$ \cms. \footnote{A value of $v_c$ as large as $\sim 10^7$
\cms\ was estimated in \citet{link09}, assuming $E_p=5$ MeV at
$\rho=10^{13}$ \gcc, for $\xi\simeq a\simeq 70$ fm. This number is a 
generous upper limit.} 

The corresponding differential angular velocity between the superfluid
and the crust is as large as $\sim 1$\ rad s$^{-1}$, but still much
less than the angular velocity of the star when the superfluid
condensed. The relative flow between the superfluid and the crust will
thus be close to or comparable the local critical velocity in regions
where there is pinning. We now examine the stability of this
differentially-rotating state.

\section{Perturbation Analysis}

The problem of the coupled dynamics of the superfluid and vortex
lattice can be studied using the hydrodynamic theory of \citet{bc83}
which accounts for vortex degrees of freedom. The local quantities of
fluid velocity, vortex density, and vortex velocity are averaged over
a length scale that is large compared to the inter-vortex spacing
$l_v$; the theory is valid for wavenumbers that satisfy $kl_v<<1$.  We
treat the superfluid as a single-component fluid at zero temperature,
and ignore dissipation in the bulk fluid and the small effects of
vortex inertia. These approximations are justified in a typical
neutron star, for which the temperature of the inner crust is much
less than the condensation temperature of the superfluid. We also
treat the crust as infinitely rigid and ignore local shear
deformations, an approximation that will be 
justified below.  The motion of the superfluid does not
couple to the electrons, so electron viscosity is not
relevant. Magnetic fields are not relevant either, as they do not
interact with the vortices of the inner crust.

We will consider only shear modes in the superfluid, so that the flow
velocity $\vbf(\rbf,t)$ is divergence-free.  The rotation axis lies
along $\hat{z}$, and $\rbf_v(\rbf,t)$ denotes the continuum vortex
displacement vector, with components in the $x-y$ plane only.  The
equations of motion in the laboratory frame are \citep{bc83}
\begin{equation}
\nabla\cdot\vbf = 0 
\end{equation}
\begin{equation}
\frac{\partial\vbf}{\partial t}+\vbf\cdot\nabla\vbf = -\nabla\mu-\nabla\phi
-\sigmabf_{el}/\rho+\fbf/\rho  
\label{sfaccel} 
\end{equation}
\begin{equation}
\rho\,\omegabf\times\left(\vbf-\frac{\partial\rbf_v}{\partial
t}\right)=-\sigmabf_{el}(\rbf_v)+\fbf. 
\label{lineforce}
\end{equation}
where $\omegabf\equiv\nabla\times\vbf$ is the vorticity due to the
existence of vortices in the fluid, $\mu$ is the
chemical potential, $\phi$ is the gravitational potential,
$\sigmabf_{el}/\rho$ is the elastic force per unit volume that arises
from bending of the vortex lattice, and $\fbf/\rho$ is the force per
unit volume exerted on the fluid by the normal matter. The elastic
force is
\begin{equation}
\sigmabf_{el}/\rho=-c_T^2\left[2\nabla_\perp(\nabla\cdot\rbf_v)-
\nabla^2_\perp\rbf_v\right]+c_V^2\frac{\partial^2 \rbf_v}{\partial z^2}
\end{equation}
where 
$\nabla_\perp$ denotes a derivative with components in the $x-y$ plane
only. Here $c_T=(\hbar\Omega/4m)^{1/2}$ is the Tkachenko wave speed
\citep{tk2a,tk2}, and $\Omega$ is the spin rate of the superfluid. 
The quantity $c^2_V=(\hbar\Omega/2m)\ln(\Omega_c/\Omega)$ is
related to wave propagation along the rotation axis;
$\Omega_c=h/(\sqrt{3}m\xi^2)$ for a triangular vortex lattice. 
For a typical neutron star rotation rate of
$\Omega=100$ rad s$^{-2}$, $c_T=0.09$ cm s$^{-1}$ and
$c_V=9\,c_T$. The areal density of vortices in the $x-y$ plane is
$l_v^{-2}=2m\Omega/h$ for a uniform vortex lattice; hence, the
requirement that $kl_v<<1$ is equivalent to $kc_T<<\Omega$.  

Eq. (\ref{lineforce}) is an expression of balance of the Magnus
force, the elastic force of the deformed vortex lattice, and the force
exerted on the fluid by the normal matter. If the vortex array is
perfectly pinned to the normal matter of the inner crust moving at
velocity $\vbf_n$, so that $\partial\rbf_v/\partial t=\vbf_n$, the
force is
\begin{equation}
\fbf=\rho\,\omegabf\times(\vbf-\vbf_n)+\sigmabf_{el}(\rbf_v). 
\end{equation}
For imperfect pinning, the Magnus force and elastic force drive 
vortex motion with respect to the normal matter. 
For imperfect pinning, the force above can be generalized as 
\begin{equation}
\fbf= 
\beta^\prime\rho\,\omegabf\times(\vbf-\vbf_n)
+\beta\rho\,\hat{\omega}\times(\omegabf\times\{\vbf-\vbf_n\})
+(1-\gamma)\,\sigmabf_{el}(\rbf_v). 
\label{fl}
\end{equation}
The first two terms of this force are present in the mutual
friction force introduced by \citet{hv56}. 
We emphasize the generality of the force law of eq. (\ref{fl}). The
first two terms represent the force exerted on the fluid by vortices
that are moving with respect to the normal matter; the first term
corresponds to the force transverse to the vortex motion, while the
second term corresponds to the force parallel to the vortex
motion. The coefficients $\alpha$ and $\beta$ can be calculated using
a specific theory of vortex mobility. The third term accounts for the
the contribution to the force that arises from local vortex bending.

If the vortex lattice is
locally undeformed ($\sigmabf_{el}=0$), the vortex velocity from
eqs. (\ref{fl}) and (\ref{lineforce}) is
\begin{equation}
\frac{\partial\rbf_v}{\partial t}=\vbf_n+\alpha\,(\vbf-\vbf_n)-
\beta\,\hat{\omega}\times(\vbf-\vbf_n), 
\label{vv}
\end{equation}
where $\alpha\equiv 1-\beta^\prime$.  Imperfect pinning, that is,
``vortex creep'', corresponds to $\alpha<<1$ and $\beta<<1$.  We refer
to $\alpha$, $\beta$, and $\gamma$ as the ``pinning coefficients''.
Perfect pinning corresponds to the limit $\alpha=\beta=\gamma=0$,
while no pinning ($\fbf=0$) corresponds to $\alpha=\gamma=1$ and
$\beta=0$.  Vortices move with a component along $\vbf-\vbf_n$, so
that $0<\alpha\le 1$. The energy dissipation rate per unit volume is
determined by $\beta$, which must be positive to give local entropy
production.

Vortex creep could be a low-drag process, with $\beta<<\alpha$, or a
high-drag process, with $\beta>>\alpha$. In
much previous work on pinning, the high-drag limit has been implicitly
assumed through the following relationship between 
$\beta$ and $\beta^\prime$: 
\begin{equation}
\beta^\prime= 1-\alpha=\frac{{\cal R}^2}{1+{\cal R}^2}={\cal R}\,\beta, 
\label{drag}
\end{equation}
where ${\cal R}$ is a dimensionless drag coefficient. In this drag
description, imperfect pinning ($\alpha<<1$, $\beta<<1$) corresponds
to ${\cal R}>>1$ so that eq. (\ref{drag}) requires
$\beta>>\alpha$. Eq. (\ref{drag}) {\em is not true in general}. The
presence of non-dissipative forces between vortices and the solid to
which they are pinned can give $\beta<<\alpha$ for $\alpha$ and
$\beta$ both small, a regime of low drag that does not follow
from eq. (\ref{drag}) for any value of ${\cal R}$ \citep{link09}. As
discussed below, it is the low-drag regime that is likely to be
realized, with vortex creep being unstable in this regime. A crucial
feature of our analysis is that we do not assume eq. (\ref{drag}).

To examine the stability of superfluid flow with imperfect pinning, we
use a local plane wave analysis in the frame rotating with the normal
matter at angular velocity $\Omegabf_n$, in which $\vbf_n=0$ and the
unperturbed flow velocity arising from spin down of the crust is
$\vbf_0$. Restricting the analysis to the regime $k\Delta R>> 1$,
where $\Delta R$ is the thickness of the inner crust, the background
flow can be taken to be uniform and the local analysis is
valid. We take the unperturbed vortex lattice to be locally undeformed
($\sigmabf_{el}=0$). The unperturbed creep
velocity in the rotating frame ($\vbf_n=0$) follows from eq. (\ref{vv}):
\begin{equation}
\frac{\partial\rbf_{v0}}{\partial t}
=\alpha\,\vbf_0-\beta\,\hat{\omega}\times\vbf_0. 
\label{vv0}
\end{equation}
Below we estimate $\partial r_{v0}/\partial t\sim 10^{-5}\,v_0$
for a typical neutron star. 
Linearizing eqs. (\ref{sfaccel}) and (\ref{lineforce}) about $\vbf_0$,
$\partial\rbf_{v0}/\partial t$, and $\sigmabf_{el}=0$, and neglecting
$\partial\rbf_{v0}/\partial t$ compared to $\vbf_0$, gives
\begin{equation}
\nabla\cdot\delta\vbf=0 \label{div} 
\end{equation}
\begin{equation}
\frac{\partial\delta\vbf}{\partial t}+\vbf_0\cdot\nabla\delta\vbf+
2\Omegabf_n\times\delta\vbf= 
-\nabla\delta\mu^\prime-\nabla\delta\phi -\sigmabf_{el}/\rho
+\delta\fbf/\rho 
\end{equation}
\begin{equation}
\rho\,
2\Omegabf_n\times\left(\delta\vbf
-\frac{\partial\delta\rbf_v}{\partial t}\right)
+\rho\,\delta\omegabf\times\vbf_0 = -\sigmabf_{el}+\delta\fbf 
\label{vortexaccel1}
\end{equation}
where $\delta$ denotes a perturbed quantity,
and $\mu^\prime\equiv\mu-\rho(\Omegabf_n\times\rbf)^2/2$. We assume 
that $\alpha$ and $\beta$ are constants. The perturbed force is then
\begin{equation}
\delta\fbf= 
(1-\alpha)\rho\,\delta\left\{\omegabf\times\vbf\right\}\nonumber 
+\beta\rho\,\delta\left\{\hat{\omega}\times(\omegabf\times\vbf)\right\}
 +(1-\gamma)\,\sigmabf_{el}(\rbf_v). 
\label{df}
\end{equation}
The vorticity appearing in this equation is the total vorticity evaluated
in the laboratory frame. For $\nabla\times\vbf_0<<2\Omegabf_n$, a good
approximation for most neutron stars, the vorticity is 
\begin{equation}
\omegabf=2{\mathbf\Omega_n}+\nabla\times\delta\vbf. 
\end{equation}
The final term in eq. (\ref{df}), associated with stress in the vortex
lattice, will turn out to be negligible for vortex creep driven by a
flow $v_0>>c_T$ and $v_0>>c_V$. 

We take the rotation axis to be $\hat{z}$, with the unperturbed
flow in the azimuthal direction, and along $\hat{x}$ at some
point. For simplicity, we restrict $\kbf$ to lie in the $x-z$ plane,
with an angle $\theta$ with respect to the rotation axis.  We further
restrict the analysis to the quadrant $0\le\theta\le\pi/2$. For shear
perturbations, $\kbf\cdot\delta\vbf=0$, that is, the velocity
perturbations in the directions $\hat{y}$ and
$\hat{e}\equiv-\cos\theta\,\hat{x}+\sin\theta\, \hat{z}$ are
orthogonal to $\kbf$.

We now Fourier transform $(\propto {\rm e}^{i{\mathbf k}\cdot{\mathbf
r}-i\sigma t})$ eqs. (\ref{div})-(\ref{df}) and take the projections
onto $\hat{y}$ and $\hat{e}$. Defining
$\sigma^\prime\equiv\sigma-kv_0\sin\theta$, $c\equiv\cos\theta$, and
$s\equiv\sin\theta$, we obtain the system of equations:
\begin{equation}
 \left[ 
\begin {array}{cccc} 
-i\sigma^\prime+2\Omega_n\beta-i(1-\alpha)kv_0s &
-2\Omega_n\alpha c &
\gamma(c_T^2s^2+c_V^2c^2)k^2 &
0 \\
-i\beta kv_0 s c +2\Omega_n\alpha c & 
-i\sigma^\prime +2\Omega_n\beta c^2 -i(1-\alpha)kv_0 s & 
0 &
-\gamma(c_T^2s^2-c_V^2c^2)k^2  \\
-i(\sigma^\prime+kv_0 s) &
0 &
0 &
i2\Omega_n\sigma^\prime/c \\
0 &
-i(\sigma^\prime + kv_0 s) &
-i2\Omega_n c\sigma^\prime &
0 \\
\end {array} 
\right] 
\left[
\begin{array}{c}
\hat{y}\cdot\delta\vbf \\
\hat{e}\cdot\delta\vbf \\
\hat{y}\cdot\rbf_v \\
\hat{e}\cdot\rbf_v \\
\end{array}
\right]=0
\label{matrix}
\end{equation}
The resulting dispersion relation is quadratic:
\begin{equation}
(\sigma^\prime)^4+a_3(\sigma^\prime)^3+a_2(\sigma^\prime)^2
+a_1\sigma^\prime+a_0=0,
\label{dr}
\end{equation}
where, in units with $\Omega_n=1$, 
\begin{equation}
a_3=2(1-\alpha)s\{kv_0\}+2i(1+c^2)\beta
\end{equation}
\begin{eqnarray}
a_2=(\alpha-1)^2 s^2 \{kv_0\}^2 &+&
\left(2i\beta s(1+c^2)-2i\alpha\beta s +\frac{1}{2}i\beta\gamma s^3
\{kc_T\}^2
  +\frac{1}{2}\beta\gamma sc^2 \{kc_V\}^2\right)\{kv_0\} \nonumber \\
&-& 4(\alpha^2+\beta^2)c^2
-\alpha\gamma s^4 \{kc_T\}^2
-\alpha\gamma c^2(1+c^2)\{kc_V\}^2
+\frac{1}{4}\gamma^2 s^4 \{kc_T\}^4
-\frac{1}{4}\gamma^2 c^4 \{kc_V\}^4
\end{eqnarray}
\begin{equation}
a_1=
\frac{1}{2}i\beta\gamma(\{kc_T\}^2 s^4 +\{kc_V\}^2 c^2s^2)\{kv_0\}^2
-\left(\alpha\gamma[\{kc_T\}^2 s^5 +\{kc_V\}^2 s c^2(1+c^2)]
-\frac{1}{2}\gamma^2 [\{kc_T\}^4 s^5 -\{kc_V\}^4 sc^4]\right)\{kv_0\}
\end{equation}
\begin{equation}
a_0=\frac{1}{4}\gamma^2(\{kc_T\}^4 s^6 - \{kc_V\}^4 s^2c^4)\{kv_0\}^2,
\end{equation}
where, $\{kv_0\}\equiv kv_0/\Omega_n$, 
$\{kc_T\}\equiv kc_T/\Omega_n$, and
$\{kc_V\}\equiv kc_V/\Omega_n$. 

\section{Wave solutions without flow}

Before turning to the full problem with non-zero $v_0$, we consider
the limit of $v_0=0$ for the two cases of zero pinning and imperfect
pinning.  The dispersion relation is quadratic
\begin{equation}
\sigma^2
+2i\beta(1+c^2)\sigma
-4c^2(\alpha^2+\beta^2)
+\gamma c_T^2 k^2 s^4\left(-\alpha+\frac{1}{4}\gamma c_T^2 k^2\right)
-\gamma c_V^2 k^2 c^2\left(\alpha (1+c^2)
+\frac{1}{4}\gamma c_V^2 k^2c^2\right)
=0, 
\label{dr0}
\end{equation}
For zero pinning force ($\alpha=\gamma=1$, $\beta=0$), 
the dispersion relation to order $c_T^2k^2$ and $c_V^2k^2$
becomes 
\begin{equation}
\sigma^2= (2\Omega_n\cos\theta)^2+c_V^2(k^2\cos^2\theta)(1+\cos^2\theta)
+c_T^2k^2\sin^4\theta, 
\label{bc}
\end{equation}
as found by \citet{bc83}. 
The fluid supports Tkachenko modes for $\theta=\pi/2$, and
axial modes (modified inertial modes) for $\theta=0$. In the limit
$c_T=c_V=0$, the system supports only ordinary inertial modes. 

The role of pinning can be seen by
considering axial modes for the case $\gamma
c_V^2k^2<<\alpha\Omega_n^2$. The solutions to eq. (\ref{dr0}) in this
limit are 
\begin{equation}
\sigma_\pm =2i\beta\,\Omega_n\pm\left(2\alpha\,\Omega_n+\frac{1}{2}
\frac{\gamma c_V^2 k^2}{\Omega_n}\right), 
\label{axial}
\end{equation}
which shows the damping effect of $\beta$. 
Pinning strongly suppresses the axial mode given by eq. (\ref{bc}),
eliminating it entirely for perfect pinning. The waves are
underdamped for $\beta<\alpha$, which defines the regime of low drag
that we will study further. 

\section{Instability}

We now show that a non-zero background flow $\vbf_0$ drives a
hydrodynamic instability if the vortices are imperfectly pinned
($\alpha<<1$, $\beta<<1$, $\gamma<<1$). We are interested in flow
velocities of $v_0\sim 10^5$ \cms. By comparison,
\begin{equation}
c_V\sim 10\, c_T \sim 10^{-5} v_0
\end{equation}
We will find that there is an instability for wavenumbers $k\gap 
\Omega_n/v_0$. The hydrodynamic limit imposes the restriction
$kc_T/\Omega_n<<1$. The regime of interest is thus, 
\begin{equation}
\Omega_n/v_0<k<<\Omega_n/c_T.
\end{equation}
We will estimate below that $\alpha\sim 10\,\beta\sim 10^{-10}$. We
assume that $\gamma$ is similarly small. 

We will not present here an analysis of the full mode structure of the
system, but focus on two low-frequency modes that appear for imperfect
pinning. We simplify the problem by proceeding 
to linear order in the small quantities $kc_T/\Omega_n$ and
$kc_V/\Omega_n$. At this level of approximation: 
\begin{equation}
a_3=2(1-\alpha)s\{kv_0\}+2i(1+c^2)\beta
\end{equation}
\begin{equation}
a_2=(\alpha-1)^2 s^2 \{kv_0\}^2 +
2i\beta s(1+c^2-\alpha) \{kv_0\}-4c^2(\alpha^2+\beta^2)
\end{equation}
\begin{equation}
a_1=a_0=0, 
\end{equation}
that is, the vortex lattice exerts no 
stresses on the fluid to first order in $c_T/v_0$ and
$c_V/v_0$. 
The dispersion relation eq. (\ref{dr}) now simplifies to:
\begin{equation}
(\sigma^\prime)^2\left\{(\sigma^\prime)^2
+2(\{1-\alpha\} kv_0s+  i\beta\{1+c^2\})\,\sigma^\prime
+(1-\alpha)^2 k^2v_0^2s^2
-4c^2(\alpha^2+\beta^2)
+2i\beta kv_0s(1+c^2-\alpha)\right\}=0.
\label{dr1}
\end{equation}
The two degenerate solutions $(\sigma^\prime)^2=0$
correspond to $\sigma=kv_0\sin\theta=\kbf\cdot\vbf_0$, the frequency
associated with translation of the wave pattern at velocity $\vbf_0$.
The other two other solutions are, switching from $\sigma^\prime$ to
$\sigma$ and restoring $\Omega_n$,
\begin{equation}
\sigma_\pm =\alpha kv_0\sin\theta-i\Omega_n(1+\cos^2\theta)\beta 
\pm \left(
4\Omega_n^2\alpha^2 \cos^2\theta
-\Omega_n^2\beta^2\sin^4\theta
-2i\alpha\beta \Omega_nkv_0\cos^2\theta\sin\theta\right)^{1/2}.
\label{solutions}
\end{equation}
For $\beta<<\alpha$ and low wavenumber $kv_0<<\Omega_n$, there are two
damped modes
\begin{equation}
\sigma_\pm\simeq \alpha(kv_0\sin\theta\pm 2\Omega_n\cos\theta)
-i\beta\left(\Omega_n\{1+\cos^2\theta\}\pm\frac{1}{2}kv_0\cos\theta\,\sin\theta
\right).
\end{equation}
Slow vortex motion has introduced two low-frequency modes to the
system. Removing pinning and drag ($\alpha=1$, $\beta=0$) and taking
$k=0$, we recover the ordinary inertial modes $\sigma_\pm =\pm
2\Omega_n\cos\theta$.

Above a critical wavenumber $k_c$, the the solution with 
eigenvalue $\sigma_-$ is unstable: 
\begin{equation}
k>k_c\equiv2\frac{\Omega_n}{v_0}\frac{(\beta^2+\alpha^2)^{1/2}}
{\alpha}
\frac{1+\cos^2\theta}{\sin\theta\cos\theta}.
\label{kc}
\end{equation}
Numerical solution of the full dispersion relation,
eq. (\ref{matrix}), for reasonable values of $c_T$, $c_V$, and $v_0$,
confirms that there are no other instabilities. The critical
wavenumber $k_c$ is minimized for $\theta=\tan^{-1}(\sqrt{2})$. For
$k>>k_c$, we have the approximate solutions
\begin{equation}
\sigma_\pm\simeq
\alpha kv_0\sin\theta\mp
i(\alpha\beta\,\Omega_nkv_0\cos^2\theta\sin\theta)^{1/2}. 
\label{highk}
\end{equation}

The instability arises from coupling
between velocity and vorticity through the first two terms of
eq. (\ref{df}).  Dissipation damps perturbations for $k<k_c$, but for
$k>k_c$ the finite vortex mobility gives rise to growing perturbations
under the Magnus force.  For $k>>k_c$, the growth rate scales as
$(\alpha\beta v_0)^{1/2}$. For $\beta<<\alpha$, $k_c$
takes a constant value, but the growth rate of the mode becomes small,
going to zero as $\beta$ goes to zero. In the highly-damped regime,
$\beta>>\alpha$, damping restricts the unstable mode to large $k$,
generally stabilizing the system. There are no unstable modes for
either $\alpha=0$ or $\beta=0$; the instability occurs only if the
vortices move with respect to the crust, both along the flow and
transverse to the flow.  

We now show that our neglect of shear deformations of the crust is a
good approximation. The modes we are studying are in the regime
$kv_0>\Omega_n$. In this limit, the dominant contribution to the shear
force per unit volume in the fluid is (see eq. \ref{df})
\begin{equation}
\delta f/\rho \sim kv_0\delta v. 
\end{equation}
Because the vortices are nearly perfectly pinned, this shear force 
creates a strain field in the solid with a shear force per unit volume
of 
\begin{equation}
\delta f_s/\rho\sim c_s^2 k^2 \delta u, 
\end{equation}
where $c_s$ is the shear speed of the solid and $\delta u$ is the
characteristic displacement. The speed of a mass element in the solid
is $\delta v_n\sim \mbox{Re($\sigma_\pm$)}\delta u$. Equating $\delta
f$ and $\delta f_s$, and using eq. (\ref{highk}) for the limit of large
$k$, gives
\begin{equation}
\frac{\delta v_n}{\delta v}\sim \alpha\left(\frac{v_0}{c_s}\right)^2. 
\end{equation}
The values $\alpha=10^{-10}$, $v_0=10^5$ \cms, and $c_s=10^8$ \cms,
give $\delta
v_n/\delta v\sim 10^{-16}$. The displacement of the solid is very
small for two reasons: i) the solid is very rigid compared to the
vortex lattice, and, ii) the vortex creep modes are of very low
frequency, proportional to $\alpha<<1$.

We have restricted the analysis to shear waves
($\nabla\cdot\delta\vbf=0$). Because these waves to not perturb the
density, we do not expect the finite compressibility of the matter to
change our results. Compressibility will introduce new modes
\citep{haskell11}, an effect that merits further study in the
context of imperfect pinning. 

\section{Estimates}

To obtain the growth rate of the instability, we now estimate the
pinning parameters $\alpha$ and $\beta$ for the vortex creep process.
To make these estimates, we regard the process of vortex creep
as consisting of two distinct states of motion for a given vortex
segment. Most of the time, the vortex segment is pinned. A small
fraction of the time, the vortex segment is translating against a drag
force to a new pinning configuration. The mutual friction force we are
using (eq. \ref{fl}) is, ignoring the small force from the vortex
lattice, 
\begin{equation}
\fbf/\rho=
\omegabf\times\vbf
-\alpha\,\omegabf\times\vbf
+\beta\, \hat{\omega}\times(\omegabf\times\vbf).
\label{fagain}
\end{equation}
This force represents the average force exerted on the neutron fluid by the
vortex array. The first term is
the Magnus force for perfect pinning, while the remaining terms give
the contribution to the force due to vortex motion. For those vortex 
segments that are unpinned and moving against drag, we take the force
to have the same form, but with different coefficients:
\begin{equation}
\fbf_0/\rho=
\omegabf\times\vbf
-\alpha_0\,\omegabf\times\vbf
+\beta_0\, \hat{\omega}\times(\omegabf\times\vbf).
\label{fmf0}
\end{equation}
An unpinned vortex segment remains unpinned for a time $t_0\sim d/v_0$,
where $d$ is the distance the segment moves before repinning. This
distance is comparable to the distance between pinning sites
\citep{leb93}, roughly ten times the unit cell size, giving
$t_0\sim 10^{-15}$ s, 
much shorter than the hydrodynamic timescales of interest. 
Suppose that at any instant, the fraction of vortex
length that is unpinned is $f_v<<1$. 
We now average $\fbf_0$ over 
a volume that contains many vortices, and
over a time long compared to $t_0$ but short compared to hydrodynamic
timescales, to obtain
\begin{equation}
\langle{\fbf_0}/\rho\rangle=
\omegabf\times\vbf
-f_v\alpha_0\,\omegabf\times\vbf
+f_v\beta_0\, \hat{\omega}\times(\omegabf\times\vbf).
\label{fave}
\end{equation}
Quantities related to the flow are unchanged by the averaging
procedure since the superfluid flow velocity is independent of whether
vortices are pinned or not.  The factors of $f_v$ in eq. (\ref{fave})
account for the fact that only the motion of the translating vortex segments
contributes to the mutual friction (see, also,
\citealt{jahanmiri06}). The value of $f_v$ is
unimportant for the following estimates.

The force of eq. (\ref{fagain}), which is appropriate for vortex
creep, must equal the average force $\langle{\fbf_0}/\rho\rangle$,
giving the following relationships: 
\begin{equation}
\alpha=f_v\,\alpha_0
\quad
\mbox{and}
\quad
\beta=f_v\,\beta_0 
\quad
\Rightarrow 
\quad
\frac{\beta}{\alpha}=\frac{\beta_0}{\alpha_0}
\label{alphabeta}
\end{equation}
We now use estimates of $\beta_0/\alpha_0$ to obtain the ratio
$\beta/\alpha$. 

The dominant drag process on unpinned vortex segments considered so
far arises from the excitation of Kelvin modes as the vortex moves
past nuclei.  Calculations of dissipation by Kelvin phonon production
on a long vortex with periodic boundary conditions for $v_0\sim 10^7$
\cms\ give typical values of $\beta_0/\alpha_0= 0.1$ and $\alpha_0\sim
1$
\citep{eb92}. Pinning occurs for $v_0\lap 10^5$ \cms, and
$\beta_0/\alpha_0$ is likely to be significantly smaller in this
velocity regime due to strong suppression of Kelvin phonon production
\citep{jones92}. Vortex creep is therefore a low-drag process if Kelvin
phonon production is the dominant dissipative mechanism. We fix
$\beta/\alpha=0.1$ for illustration in the following, which we
consider to be an upper limit; we expect typical values to be smaller.

We now estimate $\beta$. We adopt polar coordinates $(r,\phi,z)$, with
the unperturbed vorticity along $\hat{z}$ and the unperturbed flow
$\vbf_0$ along $\hat{\phi}$, and take the unperturbed flow and vortex
velocity field to be axisymmetric.  In the rotating frame, 
the unperturbed vortex velocity from eq. (\ref{vv0}) is
\begin{equation}
\frac{\partial\rbf_{v0}}{\partial t}=
\alpha\,v_0\, \hat{\phi}+\beta\, v_0\,\hat{r} 
=\hat{n}\, \frac{\partial r_{v0}}{\partial t}
\label{vv0again}
\end{equation}
where $\hat{n}$ is the average direction of vortex motion. 

For steady spin down of the star, the inner crust superfluid and the crust
are spinning down at the same rate for a local differential velocity
$v_0$. The creep velocity in this steady state is related to the
spin-down rate by \citep{alpar_etal84,leb93} 
\begin{equation}
\dot{\Omega}=-2\frac{\Omega}{r}\, 
\frac{\partial\rbf_{v0}}{\partial t}\cdot\hat{r}
=-2\frac{\Omega}{r}\, v_0\,\beta
=\dot{\Omega}_0, 
\end{equation}
where $\Omega$ is the spin rate of the superfluid, $\dot{\Omega}_0$ is
the observed spin down rate of the crust, and $r$ is approximately the
stellar radius $R$. We arrive at the estimate 
\begin{equation}
\beta=\frac{R}{4v_0t_{\rm
age}}
\simeq 10^{-11} \left(\frac{v_0}{10^5\mbox{ \cms}}\right)^{-1}
\left(\frac{t_{\rm age}}{10^4\mbox{ yr}}\right)^{-1}.
\label{ss}
\end{equation}
where $\Omega\simeq\Omega_0$ is assumed, and $t_{\rm age}\equiv
\Omega_0/2\vert\dot{\Omega}_0\vert$ is the spin-down age.
Eq. (\ref{ss}), with $\beta=0.1\alpha$, gives the 
fiducial value $\alpha\beta=10^{-21}$. For this value, we deduce
$f_v\sim (\alpha\beta/\alpha_0\beta_0)^{1/2}\sim 10^{-11}$, that is,
most of the vortex length is pinned at any instant. The 
unperturbed vortex creep speed, from eq. (\ref{vv0again}), 
is $\sim\alpha\, v_0\sim 10^{-5}$ \cms\
$<<v_0$, justifying the neglect of 
$\partial\rbf_{v0}/\partial t$ compared to $\vbf_0$ in the stability
analysis. 

We can now proceed with estimates of the instability length scale and
growth rate. 
For $\beta<\alpha$ and $\theta=\tan^{-1}(\sqrt{2})$ in eq. (\ref{kc}),
the critical wavenumber is
\begin{equation}
k_c\simeq 6\,\frac{\Omega}{v_0}=6\times 10^{-3}
\left(\frac{\Omega}{100\mbox{ rad s$^{-1}$}}\right)
\left(\frac{v_0}{10^5\mbox{ \cms}}\right)^{-1}
\mbox{ cm$^{-1}$}, 
\end{equation}
corresponding to a wavelength $\lambda=2\pi/k\simeq 10$ m. 
For
$k>>k_c$, the growth rate from eq. (\ref{highk}) is 
\begin{equation}
\frac{1}{2\pi}\,{\rm Im}(\sigma_-)\simeq 0.6\, 
\left(\frac{\alpha\beta}{10^{-21}}\right)^{1/2}
\left(\frac{\Omega}{100\mbox{ rad s$^{-1}$}}\right)^{1/2}
\left(\frac{v_0}{10^5\mbox{ \cms}}\right)^{1/2}
\left(\frac{\lambda}{1\mbox{ cm}}\right)^{-1/2}
\mbox{ yr$^{-1}$}.
\end{equation}

The hydrodynamic
treatment is restricted to $kc_T<<\Omega$. 
To estimate how high the growth rate could be, 
we consider a maximum wavenumber defined by 
$c_Tk_{\rm max}=0.1\,\Omega$, where 
$c_T\simeq 10^{-1}\, (\Omega/100\mbox{ \rads})^{1/2}$ \cms. 
The growth rate at this wavenumber, from eq. (\ref{highk}), is 
\begin{equation}
\frac{1}{2\pi}{\rm Im}[\sigma_-(k_{\rm max})]\simeq 3\,
\left(\frac{\alpha\beta}{10^{-21}}\right)^{1/2}
\left(\frac{\Omega}{100\mbox{ rad s$^{-1}$}}\right)^{3/4}
\left(\frac{v_0}{10^5\mbox{ \cms}}\right)^{1/2}
\mbox{ yr$^{-1}$}, 
\label{highsigma}
\end{equation}
For $\Omega=100$
rad s$^{-1}$, the corresponding wavenumber is $k_{\rm max}\simeq 100$
cm$^{-1}$.  Eq. (\ref{highsigma}) does not represent a physical limit,
but only the restrictions of the hydrodynamic treatment; the
instability could continue to exist also for wavenumbers in the regime
$kc_T>\Omega$.

If vortex creep is in the strongly-damped regime $\beta>>\alpha$,
contrary to the estimates here, there is still a broad window for
instability. Requiring $k_c<k_{\rm max}$ gives 
\begin{equation}
\beta<2\times 10^4\, \left(\frac{v_0}{10^5\mbox{ \cms}}\right)
\left(\frac{\Omega}{100\mbox{ rad s$^{-1}$}}\right)^{-1/2}\, \alpha, 
\end{equation}
and the star will be unstable at some wavenumber that is consistent
with the hydrodynamic regime $kc_T<<\Omega$. 

\section{Discussion and conclusions}

We have identified a dissipation-driven instability that could operate
in the fluid of the neutron star crust over length scales shorter than
$\sim 10$ m and over timescales as fast as months.  This instability
is different than other superfluid instabilities already considered in
{\em two-component} systems.  In the case of liquid helium, the
Glaberson-Donnelly instability arises when the fluid normal component
has a component of flow along the rotation axis of the system
\citep{Glaberson_etal74}. That instability occurs even if the mutual
friction is zero. A variant of the Glaberson-Donnelly instability has
been studied in the mixture of superfluid neutrons and superconducting
protons of the neutron star core, and instability was found to occur
under the assumption that the vortices are perfectly pinned to the
flux tubes that penetrate the superconducting proton fluid
\citep{gaj08a,vl08}. \citet{ga09} have identified a two-stream
instability that might occur in the neutron-proton mixture of the
core, again assuming perfect pinning of vortices to flux tubes and
neglecting magnetic stresses.  By contrast, the instability described
here exists in a {\em single-component} fluid, and occurs because the
vortices can move with respect to the solid. The component of the
motion that is transverse to the flow, the component related to
$\beta$ in eq. (\ref{fl}), though dissipative, is essential for the
instability to occur.

To illustrate the basic instability, we have taken the pinning
coefficients $\alpha$ and $\beta$ to be constants. For
thermally-activated vortex creep, these coefficients will have
exponential dependence on the velocity difference between the
superfluid and the crust \citep{alpar_etal84,leb93}. We expect that
this strong velocity dependence will significantly enhance the growth
rate of the instability. The instability of the system will be
determined by four coefficients: $\alpha(v_0)$, $\beta(v_0)$, and the
derivatives $d\alpha/dv$ and $d\beta/dv$ evaluated at $v_0$. Further
work is needed to calculate these coefficients and to incorporate them
in the stability analysis. We restricted the
analysis to wavevectors that are co-planar with the rotation axis and
the unperturbed flow; more general perturbations should be studied,
including waves with compressive components.

The flow of the inner crust superfluid could become turbulent.
If the system evolves into a state of fully-developed superfluid
turbulence with a vortex tangle, the friction force in the fluid would
be better described by an isotropic \citep{gm49} or polarized
\citep{andersson_etal07} form, rather than the anisotropic form of
eq. (\ref{fl}) that is appropriate to a regular vortex array in the
initial stages of the instability. The friction force for the tangle
could be larger or smaller than the force given by
eq. (\ref{fl}). On the one hand, the tangle has more vortex
length per unit volume to interact with the solid, which tends to
increase the force for a given value of the velocity. On the other
hand, if the vortex distribution becomes highly tangled the momentum
transfer from different regions will cancel to at least some extent,
decreasing the friction force compared to that of a straight vortex
array. If the force increases, so does the effective value of $\beta$,
and the average value of the equilibrium differential velocity will
decrease (see eq. \ref{ss}).  This effect could spell trouble for
inner-crust models of glitches. By contrast, if the friction is
decreased by turbulence, there will be more excess angular momentum in
the superfluid available to drive glitches.

The instability results from the forcing of the vortex lattice through
the solid lattice. If fully-developed turbulence results, the closest
experimental analogue might be grid turbulence that has been well
studied in superfluid helium
\citep{smith_etal93}. 
Further work is needed to determine the observational consequences of
this instability, or if it is damped by some mechanism that has been
overlooked here. We have shown elsewhere that a similar instability
occurs in the mixture of superfluid neutrons and protons of the outer
core \citep{link12b}.

\section*{acknowledgments}
We thank I. Wasserman for valuable discussions. 


\label{lastpage}

\bibliography{references}

\end{document}